\documentclass[a4paper]{jpconf}
\usepackage[]{algorithm2e}
\usepackage[]{algorithmic}

\usepackage{graphicx}
\usepackage{amsmath}
\usepackage{wrapfig }
\usepackage{eurosym}
\usepackage{chngcntr}
\usepackage{multirow}
\usepackage[utf8]{inputenc}
\usepackage[acronym,nonumberlist]{glossaries}

\makeglossaries 
\newacronym{owf}{OWF}{Offshore Wind Farm(s)}
\newacronym{wt}{WT}{Wind Turbine(s)}
\newacronym{oss}{OSS}{Offshore Substation}
\newacronym{lcoe}{LCOE}{Levelized Cost of Energy}
\newacronym{npv}{NPV}{Net Present Value}
\newacronym{dcf}{DCF}{Discounted Cash Flow}
\newacronym{ew}{EW}{Esau-Williams}
\newacronym{kr}{KR}{Kruskal}
\newacronym{pr}{PR}{Prim}
\newacronym{di}{DI}{Dijkstra}
\newacronym{vam}{VAM}{Vogel's Approximation Method}
\newacronym{ga}{GA}{Genetic Algorithm}
\newacronym{l}{L}{Length}
\newacronym{i}{I}{Investment}
\newacronym{il}{IL}{Investment plus losses}
\usepackage{tabularx,booktabs}
\usepackage{bm}

\usepackage{mathtools}

\DeclarePairedDelimiter\floor{\lfloor}{\rfloor}

\usepackage{siunitx}
\usepackage{amsmath,amssymb}
\usepackage{graphicx}
\usepackage{tikz}
\usepackage{subfigure}
\usepackage{amsmath}
\usepackage{amssymb}
\usepackage{array}
\newcolumntype{C}[1]{>{\centering\arraybackslash}p{#1}}

\usetikzlibrary{arrows,automata}\usepackage{pgfplots}
\usetikzlibrary{matrix,arrows.meta}
\usetikzlibrary{shapes.geometric}
\usetikzlibrary{shapes,arrows}
\usetikzlibrary{shadows.blur}
\usepackage{tikz,times}
\usepackage[makeroom]{cancel}
\usetikzlibrary{mindmap,backgrounds}
\usetikzlibrary{positioning}
\usepackage{float}
\usepackage{array}
\usepackage{textcomp}
\usepackage{tikz}

\newcolumntype{P}[1]{>{\centering\arraybackslash}p{#1}} 

\pgfplotstableread[row sep=\\,col sep=&]{
    interval & carT & carD & carR \\
    10   & 2.25 & -2.96  & 22.10  \\
    20   & 0.55 & -0.73  & 9.05  \\
    30   & 0.14 & -0.36  & 4.95 \\
    40   & 0    & 0      & 0 \\
    50   & 0    & 0      & 0 \\
    178  & 0    & 0      & 0 \\
    }\mydata
\pgfplotstableread[row sep=\\,col sep=&]{
    interval & det & stoch\\
    10   & 1.98 & 1050\\
    20   & 1.47 & 136.7\\
    30   & 2 & 69.56\\
    40   & 2 & 64.53\\
    50   & 1.89 & 53\\
    178  & 1.98 & 7.57\\
    }\mydatatwo
\begin{document}
\title{Closed-Loop Two-Stage Stochastic Optimization of Offshore Wind Farm Collection System}

\author{Juan-Andr\'es P\'erez-R\'ua$^{\boldsymbol{\mathsf{1}}}$, Sara Lumbreras$^{\boldsymbol{\mathsf{2}}}$, Andr\'es Ramos$^{\boldsymbol{\mathsf{2}}}$ and Nicolaos A. Cutululis$^{\boldsymbol{\mathsf{1}}}$}

\address{$^{\mathsf{1}}$DTU Wind Energy, Technical University of Denmark, Frederiksborgvej 399, 4000 Roskilde, Denmark.
$^{\mathsf{2}}$Institute for Research in Technology, Comillas Pontifical University, Calle de Alberto Aguilera 23, 28015 Madrid, Spain.}

\ead{juru@dtu.dk}
\begin{abstract}
A two-stage stochastic optimization model for the design of the closed-loop cable layout of an Offshore Wind Farm (OWF) is presented. The model consists on a Mixed Integer Linear Program (MILP) with scenario numeration incorporation to account for both wind power and {cable failure} stochasticity. The objective function supports simultaneous optimization of: (i) Initial investment (network topology and cable sizing), (ii) Total electrical power losses costs, and (iii) Reliability costs due to energy curtailment from cables failures. {The mathematical optimization program is embedded in an iterative framework called PCI (Progressive Contingency Incorporation), in order to simplify the full problem while still including its global optimum}. The applicability of the method is demonstrated by tackling a real-world instance. Results show the functionality of the tool in quantifying the economic profitability when applying stochastic optimization compared to a deterministic approach, given certain values of failure parameters. 
\end{abstract}
\section{Introduction}
Offshore Wind Farms (OWFs) are shaping up as one of the main drivers towards the transition to carbon-neutral power systems. Ambitious targets set by the European Commission see offshore wind power reaching $450$ \si{GW} by 2050 \cite{Associates2019OurFuture}. Offshore electrical systems have a sizeable weight in the capital investments, reaching 15\% of the total initial expenses \cite{Sun2012TheDevelopment}, with the power cables {being the backbone component for the Balance of Plants (BoP) value and supply chain}. Furthermore, power cables can be single points of failure, leading to strongly undesired contingency \cite{ReNEWS2017RampionFault}. Shallow waters, buried depth, seabed terrain movements \cite{Warnock2019FailureSystems}, and electro-thermal stress, are differential factors in the context of OWFs, giving rise to {higher failure rates of submarine cables compared to the reported by other offshore industries, such as oil and gas \cite{CIGRE:WorkingGroupB1.102009UpdateCables} \cite{CIGRE:WorkingGroupB1.212009Third-PartyCables}}.

OWF export cables are generally built with redundancy, as the high voltage levels and long distances increase the failure probability. Likewise, cables for collection systems may also be arranged to provide greater levels of reliability, typically resulting in a closed-loop topology. However, tailor-made models to design collection system with a closed-loop structure, using global optimization, integrated with analytical methods for reliability assessment, are not readily available in the scientific literature. Radial topology, i.e., without electrical redundancy (trees according to graph theory \cite{Perez-Rua2019ElectricalReview}) has been the most common subject of study in literature in this context, and currently represents the most frequent choice by {OWF developers}. However, with the increase in the {OWFs' capacities} and the trend of moving towards subsidy-free operating regimes, quantification of economic suitability for closed-loop or radial topologies {is} becoming essential.

{Radial topology for OWF collection systems falls into a standard class of computational problems, being classified in computational complexity as NP-Hard \cite{Fischetti2018OptimizingLosses}}. Thus, scalability is the main challenge, as state-of-the-art OWFs are in the order of hundred of Wind Turbines (WTs). Mathematical models are proposed and solved through global optimization solvers in \cite{Bauer2015TheProblem}, \cite{Fischetti2018OptimizingLosses}, \cite{Pillai2015OffshoreOptimization}, \cite{Klein2017Obstacle-awareLayouts}, and \cite{Perez-Rua2019GlobalSystems}. Nevertheless, a deterministic approach is followed given the assumption of no {cable failure} {over} the project's lifetime. 

Contrarily, studies adopting stochastic techniques are available in \cite{Banzo2011StochasticFarms}, \cite{Lumbreras2013AProblems}, and \cite{Lumbreras2013OptimalStrategies}. A Mixed Integer Quadratic Program is presented in \cite{Banzo2011StochasticFarms}, which aims to analyze the suitability of having redundancy for system components subject to failure, {by solving the full stochastic optimization program including all possible contingencies}. Model reduction implementing a Mixed Integer Linear Program along with Progressive Contingency Incorporation (PCI), and decomposition strategies is performed in  \cite{Lumbreras2013AProblems} and \cite{Lumbreras2013OptimalStrategies}, proving the ability to decrease computational resources while solving to optimality small-scale OWFs {(less than or equal to 30 WTs)}. 

The latter {papers} provide {relevant} advances {in} stochastic optimization supporting several wind power and {cable failure scenarios}. Nonetheless, the inclusion of practical engineering constraints such as non-crossing of cables, {closed-loop topology, and losses inclusion in the objective function are missing in these papers}. This gap is covered in this manuscript, where a MILP optimization program based computational tool is presented, supporting decision makers during the design stage of OWFs. An algorithmic framework is developed targeting further computational simplification, supporting an objective function combining simultaneously initial investment, total electrical power losses, and energy curtailment due to cables failures. A recourse problem {(the problem of minimizing the expected cost for energy curtailment)} is solved to assess the benefits of stochastic optimization compared to the deterministic {counterpart}.

This paper is structured as follows: In Section \ref{optmod} the optimization model is explained in {detail}; followed by Section \ref{optframe}, where the algorithmic framework is explained. Finally, a case study is performed in Section \ref{compexp}. {The} work is finalized with the conclusions in Section \ref{concl}.
\section{Optimization model}
\label{optmod}
\subsection{Graph and model representation}
\label{graphrep}
The aim of the optimization is to design a closed-loop cable layout of the collection system for an OWF, i.e., to interconnect through power cables the {$n_{\text w}$} WTs to the available {Offshore Substation (OSS)}, while providing a redundant power evacuation route.

\begin{wrapfigure}{r}{0.36\textwidth}
\includegraphics[width=1\linewidth]{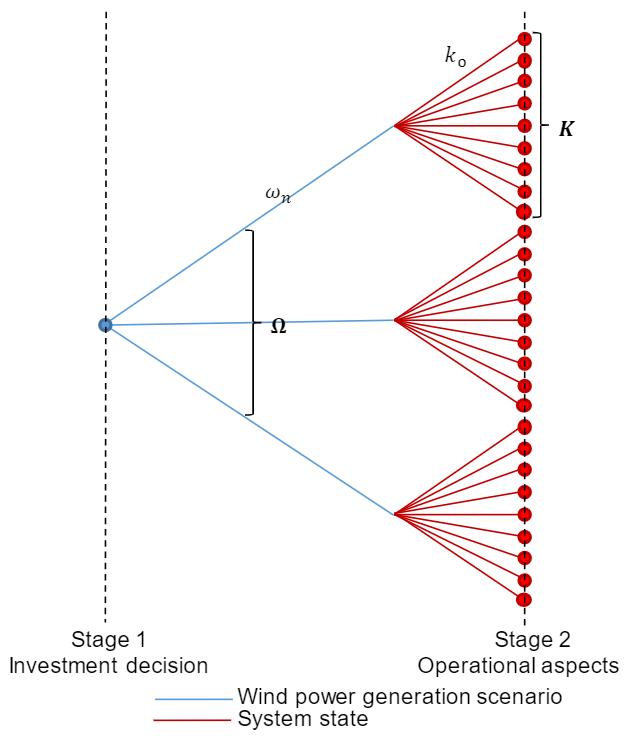}
\caption{Scenario tree, $\Upsilon$}
\label{fig:scenariotree}
\end{wrapfigure}
Let {$\boldsymbol{N_{\text w}} = \left\lbrace 2 ,\cdots,1+n_{\text w}\right\rbrace$}. {Besides, let the points set be $\boldsymbol{N} = \left\lbrace1\right\rbrace \cup \boldsymbol{N_{\text w}}$, where the element $i\in\boldsymbol{N}$, such as $i=1$ is the OSS}.

The Euclidean {distance} between the positions of the points $i$ and $j$, is {denoted by $d_{ij}$}. These inputs {are gathered} in a weighted undirected graph $G(\boldsymbol{N},\boldsymbol{E},\boldsymbol{D})$, {where $\boldsymbol{N}$ is} the vertex set, $\boldsymbol{E}$ the set of available edges arranged as a pair-set, and $\boldsymbol{D}$ the set of associated euclidean {distances} for each element $[ij] \in \boldsymbol{E}$, where $i \in \boldsymbol{N}$ $\wedge$ $j \in \boldsymbol{N}$. 

In general, $G(\boldsymbol{N},\boldsymbol{E},\boldsymbol{D})$ is a complete undirected graph{. It} may be bounded by defining uniquely those edges connecting the {$\upsilon<n_{\text w}$} closest WTs to each WT, and by the {$\sigma<n_{\text w}$} edges directly reaching the OSS {from the WTs}.

Likewise, let $\boldsymbol{T}$ be a predefined list of available cable types, and $\boldsymbol{U}$ be the set of {cable} capacities sorted in non-decreasing order as in $\boldsymbol{T}$, being measured in Amperes (\si{A}), such {that} $u_t$ is the capacity of cable $t\in\boldsymbol{T}$. Furthermore, each cable type $t\in\boldsymbol{T}$ has a cost per unit of length, {$c_{t}$ (including capital and installation costs)}, {in such a way that $\boldsymbol{U}$ and $\boldsymbol{T}$ are both comonotonic}. {The set of expenditures per meter is defined as $\boldsymbol{C}$}.

{The problem is formulated as a stochastic optimization program modelled with two stages: investment (construction) and operation. In the Figure \ref{fig:scenariotree} is presented a graphical representation of the two stages of the model. In this figure it can be seen that uncertainty is represented by means of a scenario tree ($\Upsilon$), expressing simultaneously how the stochasticity is developing over time (at the moment of the investment decision, uncertainties of the random parameters are present), the different states of the random parameters (the instances of the random process multiply in function of the generation scenarios and installed cables), and the definition of the non-anticipative decisions in the present (in real-time operation the investment decision can not be changed).} 

{The set of wind power generation scenarios is $\boldsymbol{\Omega}$ (blue lines in Figure \ref{fig:scenariotree}), while the representative system states are $\boldsymbol{K}$ (red lines in Figure Figure \ref{fig:scenariotree}). The nominal generation scenario is $\omega_{\text n}$, and the base system state ($k_{\text o}$) represents the case of no failures. The base case is therefore represented by the scenario $\left\lbrace \omega_{\text n},k_{\text o}\right\rbrace$. A wind power generation scenario $\omega$ has associated a duration time $\tau^{\omega}$ (in hours), and power magnitude $\zeta^{\omega}$ (in per unit, p.u.), and each system state $k$, a system probability $\psi^{k}$. The cost of energy in \EUR{}/\si{Ah} is denoted by $c_{\text e}$.}

{The system probability $\psi^{k}$ is calculated using a discrete Markov model to define the cables' complementary states: available, and unavailable. Through this, it is possible to calculate $\psi^{k}$ given the failure statistical parameters Mean Time Between Failures (MTBF) and Mean Time To Repair (MTTR) \cite{Calixto2016GasEngineering}. In the same way, given the low failure rates of these components a N-1 criterion must be considered in each system state \cite{Billinton1992ReliabilityTechniques}; this means that elements remaining in operation in a contingency are capable of accommodating the new operational situation, and it is very unlikely that other element would fail simultaneously.}

The first stage variables are the binary variables $x_{ij,t}$, and $y_{ij}$; where $x_{ij,t}$ is equal to one if active edge $[ij]$ ($y_{ij}=1$) uses cable type $t\in\boldsymbol{T}$. The second stage variables are the continuous variables $I_{ij}^{\omega,k}$, $\theta_i^{\omega,k}$, and $\delta^{\omega,k}_{j}$. The electrical current in edge $[ij]$ in wind power generation scenario $\omega\in\boldsymbol{\Omega}$, and system state $k\in\boldsymbol{K}$ is represented by $I_{ij}^{\omega,k}$. While the {voltage phase} at each WT busbar is $\theta_i^{\omega,k}$. The curtailed current at wind turbine $j$ in wind power generation scenario $\omega\in\boldsymbol{\Omega}$, and system state $k\in\boldsymbol{K}$ is $\delta^{\omega,k}_{j}$. Note that $\delta^{\omega,k}_{j}$ is bounded by the current generated at $j$ in the same scenario, $I^{\omega}_{j}$, where $I^{\omega}_{j}=\frac{P_n\cdot\zeta^{\omega}\cdot1000}{\sqrt{3}\cdot V_n}$, {where $P_n$ is} the nominal power of an individual WT, and $V_n$ the line-to-line nominal voltage of the system.
\subsection{Cost coefficients and objective function}
\label{costc}
\subsubsection{Neglecting total electrical power losses}
\label{displ}
The objective function in this {section} consists of a simultaneous valuation of the total initial investment plus reliability. The investment is intuitively computed as the sum of cables costs installed in each edge $[ij]$; on the other hand, reliability is quantified through the estimation of the economic losses due to cables failures, as the result of undispatched current from each WT. In this way, the objection function is formalized as:
\begin{equation} \label{eqn:objective}
\min{\overbrace{\sum\limits_{[ij] \in \boldsymbol{E}} \sum\limits_{t \in \boldsymbol{T}} c_{t} \cdot d_{ij} \cdot x_{ij,t}}^\text{Investment} + \overbrace{{c_{\text e}} \cdot \sum\limits_{i \in \boldsymbol{N_{\text w}}} \sum\limits_{\omega \in \boldsymbol{\Omega}} \sum\limits_{k \in \boldsymbol{K}} \tau^{\omega} \cdot \psi^{k} \cdot \delta^{\omega,k}_{i}}^\text{Reliability}}
\end{equation}

The sum of system states probabilities must be equal to one, {\( \sum_{k \in \boldsymbol{K}}\psi^{k}=1 \)}, given the mutually exclusive nature of the considered events ({at most} one cable is subject to {failure}, N-1 criterion).

A system state $k$ represents the failure of a single cable in an active edge $e\in\boldsymbol{E}$, therefore the system probability for the state $\psi^{k}$ is considered equal to this failure probability. {This implies that the availability probability of the other installed cables is considered to be equal to one in this scenario \cite{Banzo2011StochasticFarms}, representing a conservative approach as the value of the parameter $\psi^{k}$ is slightly overestimated (the system probability is the multiplication of each installed cable state probability)}.
\subsubsection{Considering total electrical power losses} 
\label{conpl}
Total electrical power losses are non-linear in function of the current, cable type, and total length \cite{Perez-Rua2019ElectricalReview}. The designer must try to find a proper balance between modelling fidelity and {optimization program} complexity.

{A pre-processing strategy is proposed in this manuscript in order to incorporate this factor into the objective function. This consists on calculating the quadratic total electrical power losses in advance of solving the model, by including in the objective function the costs associated to this factor.}

{The set of cable capacities in terms of number of supportable WTs is defined in the following expression}
\begin{equation} \label{eqn:capacity}
f_t =  \floor*{\frac{\sqrt{3} \cdot V_n \cdot u_t}{P_n\cdot1000}} \quad \forall t \in \boldsymbol{T}
\end{equation}

Let the new cable type set be
\begin{equation} \label{eqn:superset}
\boldsymbol{T^\prime}=\left\lbrace\underbrace{1,2,\cdots,f_1}_{t_1},\underbrace{f_1+1,\cdots,f_2}_{t_2},f_2+1,\cdots,\underbrace{f_{|\boldsymbol{T}|-1}+1,\cdots,f_{|\boldsymbol{T}|}}_{t_{|\boldsymbol{T}|}}\right\rbrace
\end{equation}

This implies that $\boldsymbol{T^\prime}$ is the discretized form of the maximum capacity $U = \max{\boldsymbol{U}}$. Note that this is translated into the creation of additional variables $x_{ij,t^\prime}: t^\prime\in\boldsymbol{T^\prime}$. Likewise, if the \textit{floor function} in (\ref{eqn:capacity}) is replaced by a \textit{decimal round down function}, and $\boldsymbol{T^\prime}$ is also discretized using {the same} decimal steps, then the number of variables will increase accordingly, to the benefit of gaining in accuracy for the cable capacities.

In $\boldsymbol{T^\prime}$ is contained the non-dominated cable sub-types from $\boldsymbol{T}$; this means that each cable sub-type $t^\prime\in\boldsymbol{T^\prime}$ is related to a cable type $t\in\boldsymbol{T}$, inheriting physical properties such as cost per meter ($c_{t}$), electrical resistance per meter ($R_{t}$), and electrical reactance per meter ($X_{t}$); see (\ref{eqn:superset}) where this relation is presented graphically. Acknowledging that the investment cost of a cable $t$ exceeds
the electrical power losses costs, then the selected cable sub-type to connect $n$ WTs will always be the cheapest (smallest) cable with sufficient capacity, rather than a bigger one with lower electrical power losses as the electrical resistance decreases with size.

As a consequence of the aforementioned, let a new {cable} capacities set be:
\begin{equation} \label{eqn:superset2}
\boldsymbol{U^\prime}=\left\lbrace1,2,\cdots,f_1,f_1+1,\cdots,f_2,f_2+1,\cdots,f_{|\boldsymbol{T}|-1}+1,\cdots,f_{|\boldsymbol{T}|}\right\rbrace \cdot \frac{P_n\cdot1000}{\sqrt{3} \cdot V_n}
\end{equation}

Let the functions $f(t^\prime)$, $g(t^\prime)$, and $h(t^\prime)$ {calculate} the cost, electrical resistance, and electrical reactance per meter for cable sub-type $t^\prime$, respectively, which are inherited from a cable type $t$. {Whereby}, the objective function for simultaneous optimization of investment, electrical losses, and reliability is:
\begin{gather} \label{eqn:objective2}
\min{\overbrace{\sum\limits_{[ij] \in \boldsymbol{E}} \sum\limits_{t^\prime \in \boldsymbol{T^\prime}} \left(f(t^\prime)+ \underbrace{3\cdot 1.5 \cdot g(t^\prime)\cdot \left(\frac{c_{\text e}}{\sqrt{3}\cdot V_n\cdot1000}\right) \cdot\sum\limits_{\omega \in \boldsymbol{\Omega}} \left(u^{\prime}_{t^\prime}\cdot\zeta^{\omega}\right)^{2} \cdot \tau^{\omega}}_{=h(t^\prime) \text{. Pre-processing for total electrical power losses}}\right) \cdot d_{ij} \cdot x_{ij,t}}^\text{Investment plus total electrical power losses}} + \\ \overbrace{c_{\text e} \cdot \sum\limits_{i \in \boldsymbol{N_{\text w}}} \sum\limits_{\omega \in \boldsymbol{\Omega}} \sum\limits_{k \in \boldsymbol{K}} \tau ^{\omega} \cdot \psi^{k} \cdot \delta^{\omega,k}_{i}}^\text{Reliability} \nonumber
\end{gather}

The factor $(3\cdot1.5)$ in (\ref{eqn:objective2}) accounts the joule, screen and armouring losses for the three-phase system. The whole term for total electrical power losses ($h(t^\prime$)) is calculated for each $t^\prime\in\boldsymbol{T^\prime}$, {before} launching the MILP program into the external solver. Therefore, the objective function is a linear weighting of the desired targets: investment, electrical losses, and reliability.

As discussed previously, one of the {tasks} of the designer is to balance out modelling fidelity and optimization program complexity. The objective function in (\ref{eqn:objective2}) is a linear function, thus the following simplifications are assumed: (i) integer discretization in (\ref{eqn:superset}) which restricts the capacity of cables, and may cause overestimation of electrical losses. This can be diminished by decimal round down, and by increasing discretization steps in (\ref{eqn:superset2}) at {the} expense of incrementing the number of variables correspondingly. (ii) Neglection of system states (cables failures) apart of the base state (no failures); however, this is the state with highest probability. (iii) Power flow estimation in a conservative fashion, i.e., overestimating the incoming power flow by neglecting the total power losses downstream. All those simplifications may impact the final layout, however their conservative nature means rather over-designing than impacting the robustness.
{\subsection{Constraints}}
\label{cons}
In case total electrical power losses are considered, the cable types set is $\boldsymbol{T^\prime}$, otherwise $\boldsymbol{T}$; same logic for $\boldsymbol{U}$/$\boldsymbol{U^\prime}$, $t$/$t^\prime$, and $u_t/u^{\prime}_{t^\prime}$. This applies for the forthcoming mathematical expressions.

The first stage constraints are first presented. These constraints are only defined by the first stage variables, i.e., the investment decision.

In case edge $[ij]$ is active in the solution, then one and only one cable type $t\in{\boldsymbol{T}}$ must be chosen, as expressed in the next equation
\begin{equation} \label{eqn:c1}
\sum\limits_{t \in {\boldsymbol{T}}} x_{ij,t} = y_{ij} \quad \forall [ij] \in \boldsymbol{E}
\end{equation}

A closed-loop (sunflower petals) collection system topology is forced through the following expression
\begin{equation} \label{eqn:c2}
\sum\limits_{\substack{j \in {\boldsymbol{N}} \\ j \neq i}} y_{ij} = 2 \quad \forall l \in \boldsymbol{N_{\text w}}: l = i \vee l = j
\end{equation}

Limiting the number of feeders (upper limit of $\phi$ feeders) connected to the OSS is carried out by means of
\begin{equation} \label{eqn:c3}
\sum\limits_{\substack{i \in {\boldsymbol{N_{\text w}}}}} y_{ij} \leq \phi \quad j = 1
\end{equation}

The set $\boldsymbol{\chi}$ stores pairs of edges $\left\lbrace[ij],[uv]\right\rbrace$, which are crossing each other. Excluding crossing edges in the solution is ensured by the simultaneous application of the next linear inequalities along with (\ref{eqn:c1})
\begin{equation} \label{eqn:c4}
y_{ij} + y_{uv} \leq 1 \quad  \forall \left\lbrace[ij],[uv]\right\rbrace \in \boldsymbol{\chi}
\end{equation}
The no-crossing cables restriction is a practical requirement in order to avoid hot-spots, and potential single-points of failure caused by overlapping cables \cite{Bauer2015TheProblem}. Constraint (\ref{eqn:c4}) exhaustively lists all combinations of crossings edges. The constraints in (\ref{eqn:c1}) ensure that no active edges are crossing or overlapping between each other. These constraints thus link the variables $y_{ij}$ and $x_{ij,t}$.

The second stage constraints are now deployed. These constraints are only defined by the second stage variables, i.e., the operational aspects. They are defined by the flow conservation, which also avoids disconnected solutions, and is considered by means of one linear equality per wind turbine as per
\begin{equation} \label{eqn:c5}
\sum\limits_{\substack{i \in {\boldsymbol{N}} \\ j \neq i}} \sum\limits_{\omega \in \boldsymbol{\Omega}} \sum\limits_{k \in
\boldsymbol{K}} I_{ji}^{\omega,k} - I_{ij}^{\omega,k} + \delta^{\omega,k}_{j} = I^{\omega}_{j} \quad \forall j \in \boldsymbol{N_{\text w}} \quad \forall \omega \in \boldsymbol{\Omega} \quad \forall k \in \boldsymbol{K}
\end{equation}

The set of tender constraints, useful to link first and second stage constraints, are lastly presented. 

A DC power flow model is applied in this manuscript, in order to calculate the power flow distribution along the resultant electrical network. This model assumes no active power losses, nominal voltage at each bar, and no reactive power flow \cite{Grainger1994PowerAnalysis}. The DC power flow is forced with the following equations
\begin{equation} \label{eqn:c6}
 I_{ij}^{\omega,k} - \frac{1000\cdot V_n \cdot (\theta_i^{\omega,k} - \theta_j^{\omega,k})}{\sqrt{3} \cdot X_{t} \cdot d_{ij}} - M \cdot (1 - x_{ij,t}) - M \cdot r_{ij}^{k} \leq 0 \quad \forall [ij] \in \boldsymbol{E} \quad t \in \boldsymbol{T} \quad \forall \omega \in \boldsymbol{\Omega} \quad \forall k \in \boldsymbol{K}
\end{equation}
\begin{equation} \label{eqn:c7}
-I_{ij}^{\omega,k} + \frac{1000\cdot V_n \cdot (\theta_i^{\omega,k} - \theta_j^{\omega,k})}{\sqrt{3} \cdot X_{t} \cdot d_{ij}} - M \cdot (1 - x_{ij,t}) - M \cdot r_{ij}^{k} \leq 0 \quad \forall [ij] \in \boldsymbol{E} \quad t \in \boldsymbol{T} \quad \forall \omega \in \boldsymbol{\Omega} \quad \forall k \in \boldsymbol{K}
\end{equation}
Where $r^k_{ij}$ is a parameter equal to one if edge $[ij]$ is failed, or zero if otherwise, $X_{t}$ is the electrical reactance per meter of cable $t$ (in case of inclusion of total electrical power losses, let $X_{t^\prime}=h(t^\prime)$), and $M$ is a big enough number to guarantee feasibility for those inactive or failed components. 

The cable capacities are not exceeded by including the next bilateral constraints.
\begin{equation} \label{eqn:c8}
\sum\limits_{t \in {\boldsymbol{T}}} u_{t} \cdot x_{ij,t} \cdot (1-r_{ij}^{k}) \geq I_{ij}^{\omega,k} \quad \forall [ij] \in \boldsymbol{E} \quad \forall \omega \in \boldsymbol{\Omega} \quad \forall k \in \boldsymbol{K}
\end{equation}
\begin{equation} \label{eqn:c9}
\sum\limits_{t \in {\boldsymbol{T}}} -u_{t} \cdot x_{ij,t} \cdot (1-r_{ij}^{k}) \leq I_{ij}^{\omega,k} \quad \forall [ij] \in \boldsymbol{E} \quad \forall \omega \in \boldsymbol{\Omega} \quad \forall k \in \boldsymbol{K}
\end{equation}
The current $I_{ij}^{\omega,k}$ may circulate either from $i$ to $j$ or viceversa.

Finally, Constraints (\ref{eqn:c10}) to (\ref{eqn:c14}) define the nature of the formulation by the variables definition, a MILP optimization program.
\begin{equation} \label{eqn:c10}
x_{ij,t} \in \left\lbrace0,1\right\rbrace \quad  \forall t \in \boldsymbol{T} \quad  \forall [ij] \in \boldsymbol{E}
\end{equation}
\begin{equation} \label{eqn:c11}
y_{ij} \in \left\lbrace0,1\right\rbrace \quad  \forall [ij] \in \boldsymbol{E}
\end{equation}
\begin{equation} \label{eqn:c12}
-0.1 \leq \theta_i^{\omega,k} \leq 0.1\quad  \forall i \in \boldsymbol{N} \quad \forall \omega \in \boldsymbol{\Omega} \quad \forall k \in \boldsymbol{K}
\end{equation}
\begin{equation} \label{eqn:c13}
 -U \leq I_{ij}^{\omega,k} \leq U  \quad  \forall [ij] \in \boldsymbol{E} \quad \forall \omega \in \boldsymbol{\Omega} \quad \forall k \in \boldsymbol{K}
\end{equation}
\begin{equation} \label{eqn:c14}
0 \leq \delta^{\omega,k}_{i} \leq I^{\omega,k}_{i} \quad  \forall i \in \boldsymbol{N_{\text w}} \quad \forall \omega \in \boldsymbol{\Omega} \quad \forall k \in \boldsymbol{K}
\end{equation}
\subsection{The stochastic optimization program}
\label{stochasticopt}
To summarize, the formulation of the MILP optimization program consists of the objective function
(\ref{eqn:objective}) or (\ref{eqn:objective2}), and the constraints defined in (\ref{eqn:c1}) - (\ref{eqn:c14}). Let this stochastic optimization program be {denoted} $P^{\boldsymbol{\Omega}, \boldsymbol{K}}$. 
{\section{Optimization framework}}
\label{optframe}
Since the two-stage variables scale-up exponentially as a function of the scenario tree size, the representative systems states must be limited \cite{Lumbreras2013AProblems}. The basic version of the stochastic optimization program presented in Section \ref{optmod} encompasses the full set $\boldsymbol{E}$; each element $[ij]$ gives place to a system state $k$ to form the system states set $\boldsymbol{K}$ (using the transformation function $\boldsymbol{K}=\Phi(\boldsymbol{E})$ which maps from edges set to system states set).

Nevertheless, the actual selected edges in a solution (i.e. a feasible point satisfying the optimality criteria) is only a subset $\boldsymbol{E^\prime}\subset\boldsymbol{E}$; let the complement set $\boldsymbol{E^{\prime\prime}}$ contain the unused elements from $\boldsymbol{E}$, and let define the subset $\boldsymbol{E^{\prime\prime\prime}}\subset\boldsymbol{E^{\prime\prime}}$.

Trough an inactive edge $[ij]$ ($y_{ij}=0$) there is no electrical current in any of the system scenarios according to (\ref{eqn:c8}) and (\ref{eqn:c9}). If there is no current through an edge, then its failure has no impact over the network power flow. This can be intuitively understood as comparing it with an open water tap which causes no spill when pipes are broken, as there is no flow of water. As a consequence, only the subset $\boldsymbol{K^\prime}\subset\boldsymbol{K}$ (related to ) is necessary and sufficient to obtain the optimum in $P^{\boldsymbol{\Omega}, \boldsymbol{K}}$.

Let the necessary and sufficient set $\boldsymbol{K^\prime}$ encompass:
\begin{equation} \label{eqn:proof1}
\boldsymbol{K^\prime}=k_{\text o} \cup \boldsymbol{K_{E^\prime}} \cup \boldsymbol{K_{E^{\prime\prime\prime}}}
\end{equation}

Where $\boldsymbol{K_{E^{\prime\prime\prime}}}$ is the system states linked to the subset of unused edges $\boldsymbol{E^{\prime\prime\prime}}$, and $\boldsymbol{K_{E^{\prime}}}$ to $\boldsymbol{E^{\prime}}$. 

The formal mathematical proof is not deployed in this article, but this particular characteristic reveals a contingency structure which can be exploited in order to simplify the full problem $P^{\boldsymbol{\Omega}, \boldsymbol{K}}$. This contingency structure opens the door for a Progressive Contingency Incorporation (PCI) strategy, aiming to find a proper set $\boldsymbol{K^\prime}$ following an iterative approach. Algorithm \ref{alg:algorithm1} is presenting the PCI implementation.

In the first line a deterministic instance of the full problem is tackled. This means considering uniquely the scenario $\left\lbrace \omega_{\text n},k_{\text o}\right\rbrace$. For this problem a valid assumption is to consider zero curtailed power. After this, the active edges of interest corresponding to the first stage optimization variables are stored as $\boldsymbol{E^\prime}$, along with the obtained solution variables in $\boldsymbol{X_{\text {ws}}}$ (where $\boldsymbol{X_{\text d}}$ and $\boldsymbol{Y_{\text d}}$ contains the solution sets corresponding to $x_{ij,t}$, and $y_{ij}$ for the deterministic case, respectively). As no previous iteration has been conducted, cumulative solution variables are unavailable ($\boldsymbol{E^{\prime}_{\text o}}$). Since the second stage variables express contingency scenarios of the components delimited by the first stage variables, the tree $\Upsilon$ uniquely considers the failure states associated to those components. For the case presented in Algorithm \ref{alg:algorithm1}, solely those feeders which satisfy the reliability level $r_\text c$, are subject to fail. 

Parameter $r_\text c$ defines the degree of connection towards the OSS, so for example, $r_\text c=1$ brings along the main feeders (rooted at $i=1$), and $r_\text c=2$ includes the last ones together with the feeders connected to the main ones, and so on for $r_\text c>2$. By means of those parameters, the model can be further relaxed for large instances.

The Progressive Contingency Incorporation routine is started at line 4. The opening step is to intersect the current active edges set $\boldsymbol{E^\prime}$, and the cumulative set $\boldsymbol{E^{\prime}_{\text o}}$. If the intersection set is equal to the current active edges $\boldsymbol{E^\prime}$, then the process is terminated, otherwise more iterations are attempted. For the former case, the algorithm is stopped, with solution $[\boldsymbol{X},\boldsymbol{Y}]$; for the latter case, the iterative process is continued to the subsequent iteration $\kappa$. Trivially, for $\kappa=1$, $\boldsymbol{A}=\emptyset$. Therefore, in line 9 the union set is obtained to update $\boldsymbol{E^{\prime}_{\text o}}$. A new instance of the main problem is solved in line 10, using the initial point $\boldsymbol{X_{\text {ws}}}$ (warm-start point), while considering the full wind power generation scenarios indicated by the user $\Omega$, and the system states related to edges cumulatively installed in all iterations, ($\boldsymbol{K^\prime}=\Phi(\boldsymbol{E^{\prime}_{\text o}})$).

When the Algorithm \ref{alg:algorithm1} converges, the scenario criterion is met: obtention of the proper set $\boldsymbol{K^\prime}$; meaning that all representative systems states have been already considered.
\begin{algorithm}[H]
\small
\caption{Progressive Contingency Incorporation (PCI) Algorithm}
\label{alg:algorithm1}
\algsetup{linenosize=\normalsize, linenodelimiter=., indent=1em}
\rule[-2ex]{0.95\linewidth}{0.5pt} 
\begin{algorithmic}[1]
\STATE{\textit{$[\boldsymbol{X_{\text d}},\boldsymbol{Y_{\text d}}]\leftarrow\arg{P^{\boldsymbol{\Omega},\boldsymbol{K^\prime}}}:\Omega=\omega_{\text n},\boldsymbol{K^\prime}=k_{\text o}$}}
\STATE{\textit{$\boldsymbol{E^\prime}\leftarrow\boldsymbol{Y_{\text d}}=\left\lbrace[ij]\right\rbrace:y_{ij}=1\quad\forall [ij]\in\boldsymbol{E}:[ij]$ satisfies reliability level $r_\text c$}}
\STATE{\textit{$\boldsymbol{E^{\prime}_{\text o}}\leftarrow\emptyset$,$\boldsymbol{X_{\text {ws}}}\leftarrow\boldsymbol{X_{\text d}}\cup\boldsymbol{Y_{\text d}}$}}
\FOR{\textit{$(\kappa=1:1:\kappa_{\text{max}})$}}
    \STATE{\textit{$\boldsymbol{A}\leftarrow \boldsymbol{E^\prime}\cap\boldsymbol{E^{\prime}_{\text o}}$}}
    \IF{\textit{$(\boldsymbol{E^\prime}==\boldsymbol{A})$}}        
        \STATE{\textit{Break}}
    \ENDIF
    \STATE{\textit{$\boldsymbol{E^{\prime}_{\text o}}\leftarrow\boldsymbol{E^\prime}\cup\boldsymbol{E^{\prime}_{\text o}}$}}    
    \STATE{\textit{$[\boldsymbol{X},\boldsymbol{Y}]\leftarrow\arg{P^{\boldsymbol{\Omega},\boldsymbol{K^\prime}}}:\Omega,\boldsymbol{K^\prime}=\Phi(\boldsymbol{E^{\prime}_{\text o}})\cup k_{\text o}$ with initial point $\boldsymbol{X_{\text {ws}}}$. $\Upsilon=\left\lbrace \Omega,\boldsymbol{K^\prime}\right\rbrace$}}
    \STATE{\textit{$\boldsymbol{E^\prime}\leftarrow\boldsymbol{Y}=\left\lbrace[ij]\right\rbrace:y_{ij}=1\quad\forall [ij]\in\boldsymbol{E}:[ij]$ satisfies reliability level $r_\text c$}}    
    \STATE{\textit{$\boldsymbol{X_{\text {ws}}}\leftarrow\boldsymbol{X}\cup\boldsymbol{Y}$}}
\ENDFOR
\end{algorithmic}
\rule[1ex]{0.95\linewidth}{0.5pt}
\end{algorithm}
{\section{Case study}}
\label{compexp}
The applicability of the method explained in Section \ref{optframe} is illustrated using Ormonde OWF \cite{VatenfallOrmondeFarm} as case study. The experiments have been carried out on an Intel Core i7-6600U CPU running at 2.50 GHz and with 16 GB of RAM. The chosen solver is IBM ILOG CPLEX Optimization Studio V12.7.1 \cite{IBM2015IBMManual}. The main input parameters are shown in Table \ref{tab:tab1}. A Mean Time To Repair (MTTR) per failure of 30 days (720 hours) is considered in this study \cite{Warnock2019FailureSystems}. The objective function (\ref{eqn:objective}) is applied in this case study as the main aim is to present the model's performance and quantitative comparison versus a deterministic version.
\vspace{-6mm}
\begin{table}[H]
\footnotesize
\caption{\label{tab:tab1}Data inputs}
\begin{center}
\begin{tabular}{c c c c c c c c c c c c c}
\br
$P_n$ & $V_n$ & Life & $c_{\text e}$ & MTTR & $\boldsymbol{U}$ & $\boldsymbol{C}$ & $n_{\text w}$ & $\upsilon$ & $\sigma$ & $\phi$ & $r_\text c$\\
\mr
5 MW & 33 kV & 30 y & 2.86 $\frac{\textup{\EUR{}}}{\si{Ah}}$ & 720 h & $\left\lbrace530,655,775\right\rbrace$ \si{A}& $\left\lbrace450,510,570\right\rbrace$ k\EUR{}/\si{km} & 30 & 6 & 10 & 4 & 1\\
\br
\end{tabular}
\end{center}
\end{table}
\vspace{-4mm}
The power magnitudes are $\zeta^{1}=1$ ($\omega_{\text n}$), $\zeta^{2}=0.5$, $\zeta^{3}=0.2$, and $\zeta^{4}=0$. The duration times account for the project's lifetime (30 years) and considering $8760$ hours per year: $\tau^{1}=65700$ hours ($\omega_{\text n}$), $\tau^{2}=91980$ hours, $\tau^{3}=91980$ hours, and $\tau^{3}=13140$ hours. A cable failure probability is calculated as per $\psi^{k}=\frac{MTTR}{MTTR+MTBT\cdot\frac{8760}{d_{ij}}}$, with $d_{ij}$ being the edge length where the component is installed. The results for this case study are obtained by varying the MTBF from $10$ to $178$, with the latter value being listed as the most critical for OWFs medium voltage cables under operation in \cite{Warnock2019FailureSystems}. Early stage in offshore projects maturity and the consequent scarcity of available data cast uncertainty over the accuracy of this parameter, implying that more critical situations can be faced in future projects. Each MTBF value represents a different stochastic problem meaning that the model is run individually. 

Quantitative assessment for the comparison between the output of the stochastic model ($[\boldsymbol{X},\boldsymbol{Y}]$), and the output of the deterministic model ($[\boldsymbol{X_{\text d}},\boldsymbol{Y_{\text d}}]$) is carried out. For the latter, a recourse problem is tackled $Q([\boldsymbol{X_{\text d}},\boldsymbol{Y_{\text d}}])$, defined as minimization of the expected costs (reliability costs) given the scenario tree ($\Upsilon$) obtained from the wind power generation scenarios $\Omega$, and the system states linked to $\boldsymbol{Y_{\text d}}$. 

For all the launched instances of MTBF an optimality gap of $0\%$ has been set up, and a reliability level of $r_\text c=1$ is assumed; this means that only those feeders connected directly to the OSS are subject to failure (the main feeders). 

\begin{figure}[H]
\centering
\centerline{\subfigure[Deterministic designed layout]{\includegraphics[scale = 0.30]{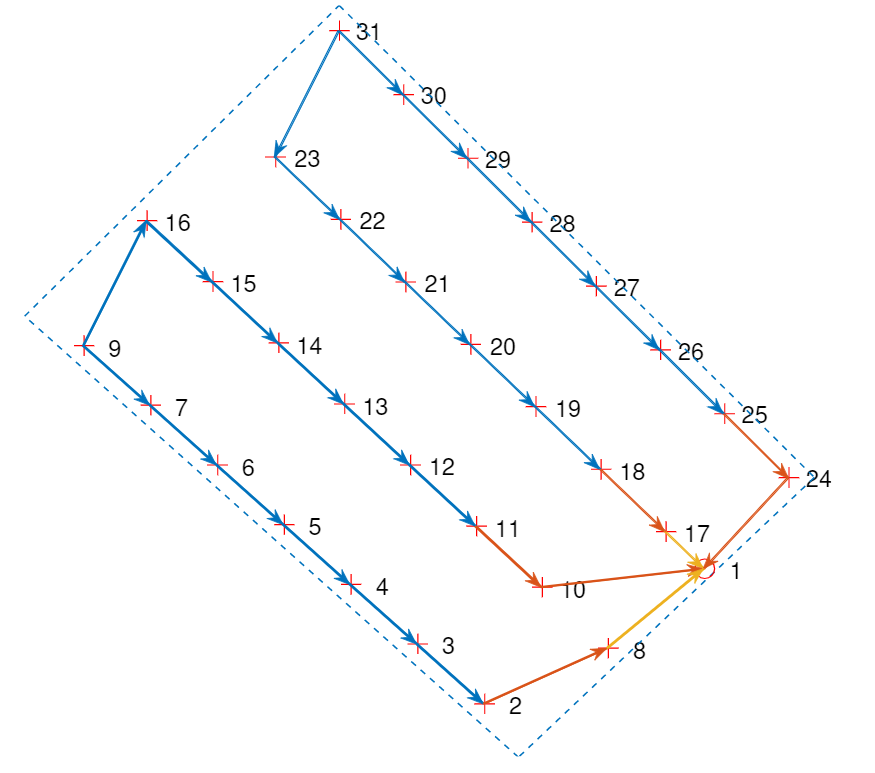}}
\subfigure[Stochastic designed layout]{\includegraphics[scale = 0.30]{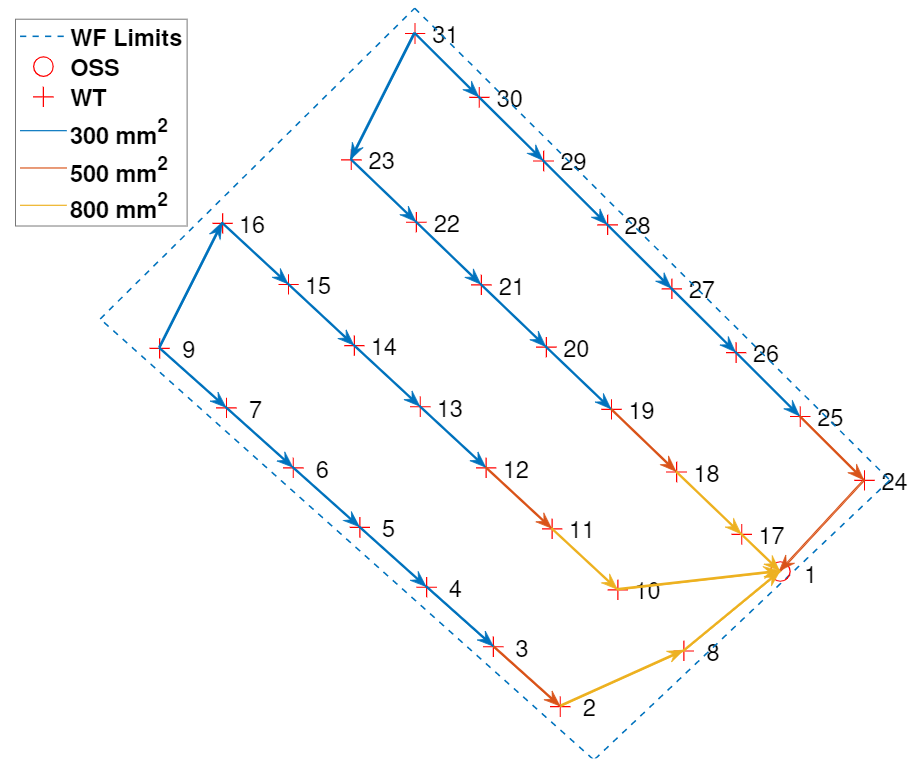}}}
\caption{Designed cables layouts for MTBF=10}
\label{fig:designed}
\end{figure}

In-detail cable layout results for a MTBF of 10 are shown in Figure \ref{fig:designed}. Due to the rather straightforward layout of the wind farm, the main difference between the deterministic and stochastic technique is the use of larger cables on the connections close to the OSS. For OWFs with more wind turbines and/or more irregular layout, the changes would likely expand to the connections between the wind turbines. 

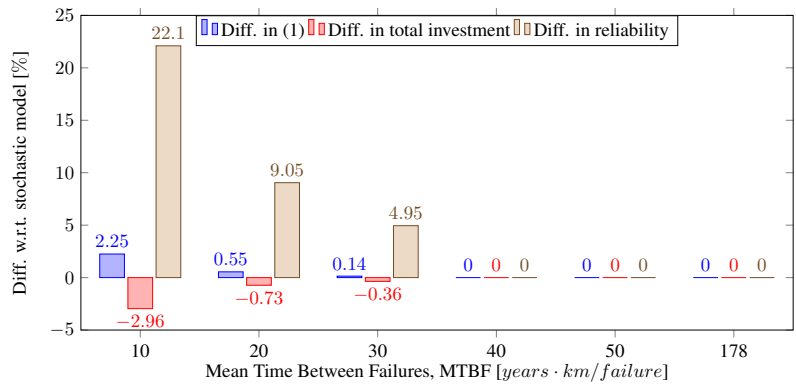
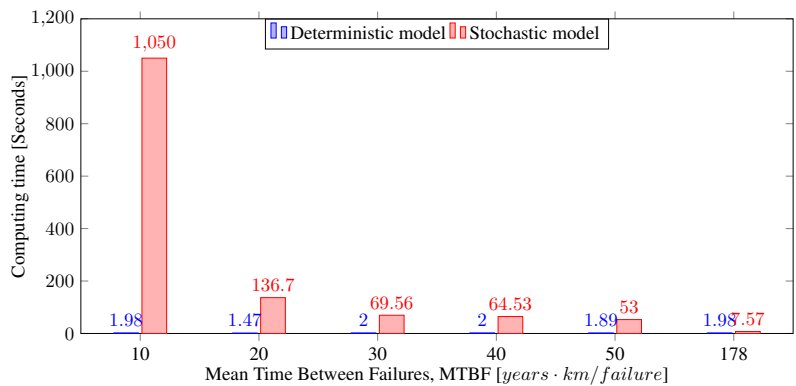
\begin{figure}[H]
\centering
\subfigure [Economic comparison. (Positive percentages mean savings using stochastic optimization)] {\begin{tikzpicture}[scale = 0.65]
    \begin{axis}[
            ybar,
            bar width=.5cm,
            width=\textwidth,
            height=.5\textwidth,
            legend style={at={(0.5,1)},
                anchor=north,legend columns=-1},
            symbolic x coords={10,20,30,40,50,178},
            xtick=data,
            nodes near coords,
            nodes near coords align={vertical},
            ymin=-5,ymax=25,
            ylabel={Diff. w.r.t. stochastic model [$\%$]},
            xlabel={Mean Time Between Failures, MTBF [$years\cdot km/failure$]},
        ]
        \addplot table[x=interval,y=carT]{\mydata};
        \addplot table[x=interval,y=carD]{\mydata};
        \addplot table[x=interval,y=carR]{\mydata};
        \legend{Diff. in (\ref{eqn:objective}), Diff. in total investment, Diff. in reliability}
    \end{axis}
\end{tikzpicture}
\label{fig:results1}
}\\
\subfigure [Computing time comparison] {\begin{tikzpicture}[scale = 0.65]
\centering
    \begin{axis}[
            ybar,
            bar width=.5cm,
            width=\textwidth,
            height=.5\textwidth,
            legend style={at={(0.5,1)},
                anchor=north,legend columns=-1},
            symbolic x coords={10,20,30,40,50,178},
            xtick=data,
            nodes near coords,
            nodes near coords align={vertical},
            ymin=0,ymax=1200,
            ylabel={Computing time [Seconds]},
            xlabel={Mean Time Between Failures, MTBF [$years\cdot km/failure$]},
        ]
        \addplot table[x=interval,y=det]{\mydatatwo};
        \addplot table[x=interval,y=stoch]{\mydatatwo};
        \legend{Deterministic model, Stochastic model}
    \end{axis}
\end{tikzpicture}
\label{fig:results2}
} 
\caption{Deterministic vs Stochastic Model for Closed-Loop Structure}
\label{fig:resultsALL}
\end{figure}

The economic comparison between deterministic and probabilistic model (see Figure \ref{fig:results1}), indicates that for MTBF values inferior to $30$, the stochastic model provides a cheaper system in overall terms; this is achieved as reliability costs are lower at the expense of more costly cable infrastructure (in this particular case only due to cable sizing). The values in Figure \ref{fig:results1} are expressed as the percentage difference between the deterministic and stochastic model relative to the latter. The associated reliability costs for the deterministic designed layout are calculated through the aforementioned recourse problem; a DC power flow and undistpached energy are the main results for this particular task.

Conversely, beyond MTBF=$30$, the failures probabilities drop considerably, resulting in an equal weight for each cost unit (initial investment and reliability costs) in the objective function of the deterministic model. This basically means that the reliability costs become trivial, and hence the focus is merely on the investment costs reduction. To compare the overall costs, the deterministic layout solution is run with the scenarios analyzed in the stochastic.

For large enough values of  $r_\text c$, for instance such that all edges are covered, one could expect that the MTBF break-even point would move further right, that is, stochastic programming would provide overall more economic benefits for less frequent failures rates.
\vspace{-6mm}
Regarding computing time, in Figure \ref{fig:results2}, it is noticeable that the stochastic model for each MTBF instance is more complex computationally, as for instance, for a value of 10 the computing time is 530 times larger than the deterministic version. Likewise, the larger the MTBF, the more simplified the model becomes as the reliability costs share are decreasing. The deterministic cases on the other hand are independent to the failures rate and converge in couple of seconds.
\section{Conclusions}
\label{concl}
A stochastic model to quantify the economic suitability of building closed-loop collection system for OWFs is introduced in this manuscript. The objective function allows for a simultaneous consideration of initial investment, total electrical power losses costs, and reliability costs. The focus of this work has been to describe the proposed model, while presenting its application in a case study.

Results of this article point out that in function of failure parameters, network topologies with redundant power corridors may bring along significant cost benefits. The stochastic model presents a complex mathematical structure impacting considerably the required computational resources. Lastly, the contingency structure of the problem has been exploited analytically in order to simplify its complexity.

Future work will present the application of this method for comparing quantitatively different network topologies, such as closed-loop and radial systems.
\section*{References}
\bibliographystyle{iopart-num}
\bibliography{references2}
\end{document}